\documentclass[sigconf,authorversion,authoryear]{acmart}

\usepackage{caption}
\usepackage{subcaption}
\usepackage{tabularx}

\AtBeginDocument{%
  \providecommand\BibTeX{{%
    \normalfont B\kern-0.5em{\scshape i\kern-0.25em b}\kern-0.8em\TeX}}}

\setcopyright{rightsretained}
\acmConference[SIGIR eCom'20]{ACM SIGIR Workshop on eCommerce}{July 30, 2020}{Virtual Event, China}
\acmYear{2020}
\copyrightyear{2020}
\makeatletter
\renewcommand\@formatdoi[1]{\ignorespaces}
\makeatother
\acmISBN{}

\begin{document}

\title{Counterfactual Learning to Rank using Heterogeneous Treatment Effect Estimation}

\author{Mucun Tian, Chun Guo, Vito Ostuni, Zhen Zhu}
\affiliation{%
  \institution{Pandora Media LLC}
  \streetaddress{2100 Franklin St}
  \city{Oakland}
  \state{CA}
  \postcode{94612}
}
\email{{mtian, cguo, vostuni, zzhu}@pandora.com}

\renewcommand{\shortauthors}{Tian, Guo, Ostuni, and Zhu}

\begin{abstract}
Learning-to-Rank (LTR) models trained from implicit feedback (e.g. clicks) suffer from inherent biases. A well-known one is the position bias --- documents in top positions are more likely to receive clicks due in part to their position advantages. To unbiasedly learn to rank, existing counterfactual frameworks first estimate the propensity (probability) of missing clicks with intervention data from a small portion of search traffic, and then use inverse propensity score (IPS) to debias LTR algorithms on the whole data set. These approaches often assume the propensity only depends on the position of the document, which may cause high estimation variance in applications where the search context (e.g. query, user) varies frequently. While context-dependent propensity models reduce variance, accurate estimations may require randomization or intervention on a large amount of traffic, which may not be realistic in real-world systems, especially for long tail queries. In this work, we employ heterogeneous treatment effect estimation techniques to estimate position bias when intervention click data is limited. We then use such estimations to debias the observed click distribution and re-draw a new de-biased data set, which can be used for any LTR algorithms. We conduct simulations with varying experiment conditions and show the effectiveness of the proposed method in regimes with long tail queries and sparse clicks.
\end{abstract}




\maketitle

\section{Introduction}
Learning-to-rank (LTR) models have been widely used in information retrieval and recommender systems. These models are often trained in the offline setting with implicit feedback (e.g. clicks) collected from production systems. While 
implicit feedback is an attractive training source (e.g. abundant, privacy preserving), its inherent biases hinder the effectiveness of learning-to-rank \citep{Joachims07}. One such bias is the position bias. To mitigate position bias, traditional approaches have gone into modeling the bias-aware relevance \citep{Joachims02, chuklin15, Craswell08}. However, accurately inferring individual relevance requires each query-document pair repeating multiple times at multiple positions, which is not realistic in many search systems. Instead of modeling individual relevance, recent counterfactual frameworks \citep{Wang16, Joachims17, Agarwal19, Fang19} attempt to estimate the examination probability under Position-Based Model (PBM) \citep{Richardson07} and use the estimation as inverse propensity score (IPS) to weight pairwise or listwise ranking. While IPS weighting provides an unbiased LTR under PBM, it has several limitations.

First, all the existing approaches follow a direct method to estimate propensities, which requires the same set of query-document pairs appearing in at least two different positions. This can be implemented by randomizing top-$N$ \citep{Wang16}, swapping pairs \citep{Joachims17, Wang18}, or integrating multiple loggers \citep{Agarwal19, Fang19}. For long tail queries, however, we rarely observe their intervention counterparts even with randomization, causing biased propensity estimations on these long tail queries. This could potentially break the unbiasedness of IPS weighted LTR. Second, the examination propensity can only be estimated on observed clicks when using the PBM model. In other words, we cannot infer from un-clicked documents whether these are irrelevant or not examined. Therefore, applying IPS into pointwise learning is not as effective as pairwise or listwise ones \citep{Wang18}. Third, IPS only weights clicked documents in the empirical loss function. Novel items (e.g. new music releases, products) for which we haven't observed any clicks, due to the lack of exposure to users, are treated as negative examples under IPS weighted LTR. This undermines LTR's ability to promote novel documents.

In this paper, we employ heterogeneous treatment effect (HTE) estimation methods \citep{Athey19, Kunzel19} to address these limitations. We first estimate causal effects of click probabilities between two positions. Based on these estimations, we then debias click distributions of observational data and draw clicks for unbiased LTR. Finally, we evaluate the effectiveness of the proposed method under varying experiment conditions using semi-synthetic data simulated from the Microsoft Learning to Rank dataset \citep{Qin_Liu13}. The objective of this work is twofold: \romannumeral 1) compare the proposed heterogeneous treatment effect methodology to existing ones, \romannumeral 2) evaluate the effectiveness of the proposed method on long tail queries.

Estimating heterogeneous treatment effect does not require intervention data, nor the click model. Instead, it utilizes "collaborative information" in the feature space, allowing the position bias estimation for long tail queries. Under the unconfoundedness assumption \citep{Rosenbaum83}, the estimator is unbiased for any given context \citep{Athey19}, so drawing from debiased click distributions can provide reliable clicks for documents with unknown relevance information in observational data. 
\section{Related Work}
There are several lines of research on estimating and/or debiasing click data for LTR. One approach is to \emph{infer relevance with heuristics or modeling}, including \emph{SkipAbove} \citep{Joachims05}, \emph{Position-Based Model} (PBM) \citep{Richardson07}, the \emph{Cascade} Model \citep{Craswell08}, and other extensions \citep{chuklin15}. These approaches attempt to derive the absolute or relative relevance by taking into account users' search behavior. While relative relevance is found to be more accurate on average, LTR trained from this relevance is likely to reverse the presented order without additional heuristics \citep{Joachims07, Joachims17}.

Another approach is \emph{Online Learning} \citep{Yue09, Schuth16}. Online learning is robust to bias and noise, but it learns from randomization data. This can hurt users' experience during the initial deployment stage \citep{Jagerman19}.

\emph{Counterfactual LTR frameworks} \citep{Wang16, Joachims17, Agarwal19-sigir} seek to estimate how likely a document is to be examined and use the inverse propensity score (IPS) to weight pairwise or listwise LTR. Counterfactual LTR does not need randomization in the learning process and is proven to be unbiased under PBM \citep{Joachims17}. However, it's sensitive to selection bias and noise \citep{Jagerman19}. Some attempts were made to keep the IPS estimator "doubly robust", such as adding an imputation term (regression on the complete data) \citep{su19}, inclusion propensity (the propensity of a new document being exposed to users) \citep{Carterette18}, or noise-aware parameters \citep{Agarwal19-www}. The robustness of these estimators rely on the accurate imputation or noise modeling. These approaches also often assume the examination propensity only depends on the position. Several techniques have been proposed to estimate the context-dependent propensity \citep{Wang16, Chandar18, Fang19}, yet all of these require intervention data. There is another line of work that jointly estimates the relevance and the position bias \citep{Wang18, Ai18, Hu19} on observational data. But coupling the relevance and the bias together without controlling for either one of them calls into question the unbiasedness of the estimator \citep{Agarwal19, Fang19}.

Recent advances in \emph{heterogeneous treatment effect estimation} provide promising techniques for identifying individual treatment effect in observational studies. This line of work follows the potential outcomes framework \citep{rubin74, imbens_rubin15} to estimate the treatment effect using \citet{Robinson88} transformation under the unconfoundedness assumption \citep{Rosenbaum83}. \citet{Kunzel19} introduced meta-learners that indirectly predict the heterogeneous treatment effect using imputation on unobserved outcomes. \citet{Athey_Imbens16} proposed recursive partitioning --- namely \emph{causal trees} --- to assess heterogeneity in the treatment effect. A following work by \citet{Wager_Athey18} developed \emph{causal forests} to consistently estimate the true treatment effect. These tree-based methods require manually-designed criteria for parameter tuning due to the fundamental problem of causal inference --- we observed an individual either in treatment or control group, but not both. \citet{nie17} proposed \emph{R-learner} that separates confounding factors from the treatment effect estimator, enabling the traditional cross-validation for goodness-of-fit. This motivates \emph{generalized random forests}'s \citep{Athey19} parameter tuning.
\section{Methods}
In this section, we briefly revisit the existing counterfactual LTR framework, then introduce the heterogeneous treatment effect estimation and describe our experiment protocol.

\subsection{Counterfactual LTR}
\label{sec: 3.1}
Counterfactual LTR frameworks \citep{Wang16, Joachims17, Agarwal19-sigir} assume that documents at higher positions are more likely to be examined by a user than ones at lower positions. Therefore, observed clicks are missing with certain propensities (probabilities) for documents at position $k$. Given these propensities, we can use inverse propensity score (IPS) technique to weight the positive examples for unbiased LTR.
\subsubsection{Position-Based Propensity Estimation}\label{sec:3.1.1}
The propensity is often unknown in practice. To estimate it, existing literature follows the position-based model (PBM) \citep{Richardson07}, which assumes the observed click $C \in \{0, 1\}$ depends on the examination $E \in \{0, 1\}$ and the relevance $R \in \{0, 1\}$ in the following way,
$$P(C = 1 | X=x, P=k) \coloneqq P(E = 1 | P=k) \cdot P(R = 1 | X=x),$$
where $X \in \mathbb{R}^t$ is the feature vector that encodes the query, user and document, and $P \in \{1, ..., K\}$ is the document position. This assumes the examination $E$ only depends on the position. We can also relax the examination to be context-dependent as 
$$P(C = 1 | X=x, P=k) \coloneqq P(E=1 | X=x, P=k) \cdot P(R=1 | X=x).$$
We then can fit the examination $\hat{f}_p(x, k)$ and the average relevance $\hat{f}_r(x)$ models with the intervention data to consistently estimate the propensity by minimizing the cross-entropy loss \citep{Wang18, Fang19},
\begin{equation}
\begin{split}
    \hat{f_p}(\cdot), \hat{f_r}(\cdot) \coloneqq \arg\min_{f_p, f_r}-\bigg\{ & \sum_i^N \sum_k^K \Big[ y_k^i \log \Big(f_p(x^i, k)\cdot f_r(x^i)\Big)\\
    & + (1-y_k^i) \log \Big(1 - f_p(x^i, k) \cdot f_r(x^i) \Big) \Big] \bigg\}, 
\end{split}
\end{equation}
where $i = 1, ..., N$ represents a unique tuple of (query, document, position) and $y_k^i$ is the click rate of a unique query-document pair at position $k$. 

\subsubsection{IPS Weighted LTR}\label{sec: 3.1.2}
IPS weighting is found to be effective to pairwise or listwise LTR \citep{Wang18}. In the IPS weighted pairwise LTR setting, we have $N$ of lists with size $K$, and we want to learn a score function $\hat{f}(\cdot)$ from the following loss,
$$\hat{f}(\cdot) \coloneqq \arg\min_f \bigg\{\frac{1}{N} \sum_i^N \sum_k^K \frac{c_k^i}{p_k^i} \mathcal{L}(f(x_1^i), ..., f(x_K^i))\bigg\},$$
where $c_k^i \in \{0, 1\}$ is the click of the $k$th document in a ranking list, $p_k^i$ is the propensity score at position $k$, and $\mathcal{L}(\cdot)$ is the pairwise loss that approximates or bounds to a ranking metric (e.g. DCG, Relevance Rank) \citep{Joachims17, Agarwal19-sigir}.  

\subsubsection{Propensity Model Implementation} We select the contextual-dependent position-based model (CPBM) \citep{Fang19} as the contextual propensity model\footnote{This was implemented using Tensorflow \citep{tensorflow2015-whitepaper}.}. Specifically, the examination model is implemented with a 3-layer neural network; the input $X \in \mathbb{R}^t$ corresponds to the context feature with size $t$, and the output $f_p(x, k) \in \mathbb{R}^K$ corresponds to the top-$K$ positions to be estimated. The average relevance is also modeled by a 3-layer neural network; the input $X \in \mathbb{R}^t$ corresponds to the context feature with size $t$, and the output $f_r(x, k, k') \in \mathbb{R}^{K \times K}$ corresponds to the intervention sets for the $K \times K$ position pairs. An intervention set is composed of documents that appear at at least two different positions. To take into account the fact that the average relevance given the context $X=x$ for intervention documents at positions $(k, k')$ is equal to the average relevance at positions $(k', k)$, the output layer is the arithmetic mean of the previous output and its transpose, making the final output a symmetric matrix (see \citep{Fang19} for the details).

\begin{figure*}[tb]
\centering
\begin{subfigure}[b]{\textwidth}
 \centering
 \includegraphics[width=\textwidth]{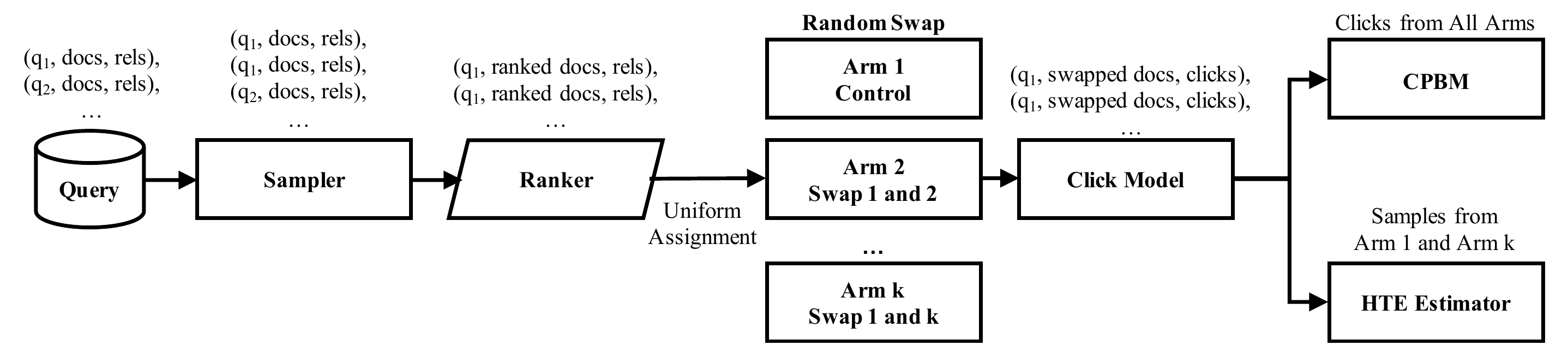}
 \caption{Position bias estimations.}
 \label{fig:pos-bias-estimations}
\end{subfigure} \\
\begin{subfigure}[b]{\textwidth}
 \centering
 \includegraphics[width=\textwidth]{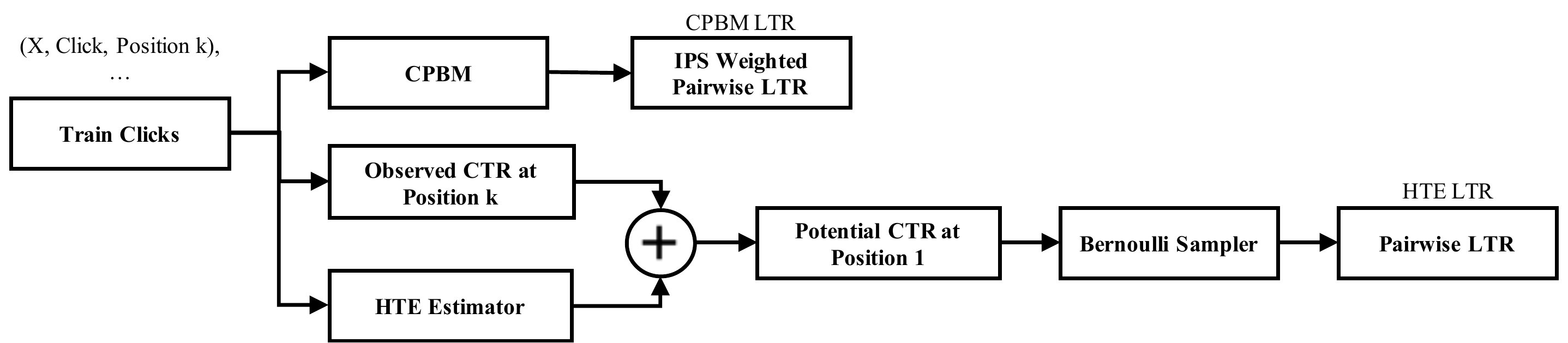}
 \subcaption{Unbiased learning-to-rank.}
 \label{fig:ltr-diagram}
\end{subfigure}
\caption{Simulation architecture.}
\label{fig:simulation-diagram}
\end{figure*}

\subsection{Heterogeneous Treatment Effect Estimation}
\label{sec: 3.2}
Another way to achieve unbiased LTR is to place every document to the first position and collect the data for offline training. This is obviously impractical, so we seek to unbiasedly estimate the conditional incremental effect --- how much is the increase of the click probability if a document would have been in the first position had it been in position $k, k \neq 1$. With the estimation, we can compensate the click probability of the document at position $k$ during the offline training.
\subsubsection{Heterogeneous Treatment Effect Estimation}
To estimate the conditional incremental effect, we employ the potential outcome framework \citep{rubin74} to formulate this problem. In LTR, we observe $N$ of i.i.d. examples $(X^{i}, Y^{i}, P^{i}), i = 1, ..., N$, where $X^{i} \in \mathbb{R}^t$ is query-document feature, $Y^{i} \in \mathbb{R}$ is the observed outcome (e.g. click, grade), and $P^{i} \in \{0, 1\}$ is the treatment variable, indicating whether a document is observed in position $k$ ($k \neq 1$) or $1$; $P^{i}=1$, if the position is $1$; $P^{i}=0$, if the position is $k$. We assume there are potential outcomes $\{Y(1), Y(0)\}$, corresponding to the treatment or control group, so $Y^{i}=Y^{i}(1)$, if $P^{i}=1$; otherwise, $Y^i=Y^{i}(0)$. We then want to estimate the conditional average treatment effect (CATE) between position $1$ and $k$,
$$\tau^{*}_k (x) \coloneqq \mathop{\mathbb{E}} [Y(1) - Y(0) \mid X = x].$$

To estimate CATE, we assume unconfoundedness given any specific context \citep{Rosenbaum83},
\begin{equation}\label{eq: unconfoundedness}
\{Y(1), Y(0)\} \perp P \mid X=x.
\end{equation}

Intuitively, this assumes data points surrounding a specific context, $X=x$, are missing at random so that we can estimate $\tau^*_k(x)$ without bias.\footnote{In practice, this assumption can be met by conducting randomization experiments. For example, we map each unique query to a random seed, and then we randomly shuffle the top-$K$ list based on this random seed. In this way, the ranking is still deterministic, but documents' positions now are independent of clicks they will receive. This reduces the harm to the user experience compared to full randomization.}

\subsubsection{Click Distribution Correction and Data Resampling} \label{3.2.2}
With the estimator $\hat{\tau}_k(x)$ in hand, we compute a potential click rate at position $1$ for each unique query-document pair ($X^i=x$) observed at position $k$ by
\begin{equation}\label{eq:click_correction}
    \hat{CTR}_1(X^i) = CTR^{obs}_k(X^i) + \hat{\tau}_k(X^i),
\end{equation}
where $CTR^{obs}_k(X^i)$ is the observed click rate of query-document pair with feature $X^i$ at position $k$.

Based on \eqref{eq:click_correction}, we re-draw $N^{obs}$ clicks/non-clicks for each unique query-document pair ($X^i=x$) from

$$\hat{C}_{X^i} \sim \mathrm{Bernoulli}(\theta^i), \; \theta^i = \hat{CTR}_1(X^i),$$
where $N^{obs}$ is the number of query-document pair $(X^i=x)$ observed in the training data. We truncate $\theta^i$ to the range $[0, 1]$ before sampling clicks.

In this way, we reconstruct clicks as if we would have been putting each document in the first position. A nice property of our approach is that relevant documents without any observed clicks due to position bias can now become positive training examples after the correction and resampling. This is particularly useful for long tail queries and long results lists. However, this is not achievable using existing counterfactual framework since IPS only weights the clicked documents.

\subsubsection{Heterogeneous Treatment Effect Implementation}
In the following we describe the two methods we used to estimate the heterogeneous treatment effect $\tau_k(x)$: \emph{causal forests} \citep{Wager_Athey18, Athey19} and \emph{X-Learner} \citep{Kunzel19}. 

Causal forests \citep{Wager_Athey18} estimates the heterogeneous treatment effect at each leaf node $L$ by
\begin{equation}
\label{eq: causal-forests}
    \begin{split}
        \hat{\tau}_k(x)= & \frac{1}{|\{i:P^i=1, X^i \in L\}|}\sum_{\{i:P^i=1, X^i \in L\}}Y^i \\
        &- \frac{1}{|\{i:P^i=0, X^i \in L\}|}\sum_{\{i:P^i=0, X^i \in L\}}Y^i.
    \end{split}
\end{equation}
Causal forests \citep{Wager_Athey18} recursively solves \eqref{eq: causal-forests} and selects a cut in the feature space that maximizes the difference of $\hat{\tau}_k(x)$ between two child nodes. To speed up the tree building process, \citep{Athey19} uses the gradient method to optimize a linear approximation of the difference (see \citep{Athey19} for the details). We train $K-1$ causal forests to estimate the bias in top-$K$ positions.

X-Learner \citep{Kunzel19} estimates the heterogeneous treatment effect by fitting models on the imputed outcomes. It has three steps:
\begin{enumerate}
    \item Fit two regression models, $\hat{\mu}_0(x)$ and $\hat{\mu}_1(x)$, to estimate the average outcomes of the treatment and control group, $\mu_0(x)=\mathbb{E}[Y(0)|X=x]$ and $\mu_1(x)=\mathbb{E}[Y(1)|X=x]$, respectively.
    \item Impute the individual treatment effects in the treatment and control group by $\tilde{D}^i_0 \coloneqq \hat{\mu}_1(X^i_0) - Y^i_0$ and $\tilde{D}^i_1 \coloneqq Y^i_1 - \hat{\mu}_0(X^i_1)$ and fit two regression models, $\hat{\tau}^0_k(x)$ and $\hat{\tau}^1_k(x)$, to estimate the imputed treatment effects, $\tau^0_k(x)=\mathbb{E}[\tilde{D}_0|X=x]$ and $\tau^1_k(x)=\mathbb{E}[\tilde{D}_1|X=x]$, respectively.
    \item Estimate the heterogeneous treatment effect by $$\hat{\tau}_k(x)=g(x)\hat{\tau}^0_k(x) + (1 - g(x))\hat{\tau}^1_k(x),$$ where $g(x) \in [0, 1]$ is often set to the propensity of treatment assignment, that is $g(x) = P(P=1 | X=x)$.
\end{enumerate}

Similarly, we train $K-1$ X-Learners for top-$K$ positions using Causal ML \citep{chen2020causalml}. We select the tree boosting method \citep{Chen_Guestrin16} as the base regression models.

\subsection{Simulation Protocol}
To evaluate the accuracy of position bias estimations and the ranking effectiveness, we simulate the entire search system, intervention experiments, and model training and evaluation. Figure~\ref{fig:simulation-diagram} shows our simulation architecture. We begin by generating intervention clicks and estimating position bias (Figure~\ref{fig:pos-bias-estimations}), and then apply bias estimations to unbiased learning to rank (Figure~\ref{fig:ltr-diagram}). In the following we detail the different steps and components of the simulation architecture.

\subsubsection{Query Sampler}
We use the Microsoft Learning-to-rank data set \citep{Qin_Liu13} as the query corpus since it provides the real-world context features and human annotated $5$-grade relevance, $rel \in \{0, 1, 2, 3, 4\}$. The data set contains 31K unique queries and their corresponding candidate documents, and it is split into train (60\%), validation (20\%), and test sets (20\%). To simulate the long tail and popular search queries, we draw uniformly from the query corpus with different sample sizes. We also train a linear pairwise ranking with $1\%$ of the train and validation queries to simulate the production ranker. For each incoming query, the production ranker outputs a ranked list for further experiments.

\subsubsection{Intervention Simulation}
Recent work collects intervention clicks from randomization \citep{Wang16, Joachims17, Wang18} or multiple loggers \citep{Agarwal19, Fang19}. For the purpose of comparison between the proposed method with the existing propensity estimation methods, we adopt an intervention simulation that is similar to the randomized swap \citep{Joachims17, Wang18}. We randomly assign each ranked query list to one of $K$ arms to create intervention rankings for position bias estimations. In the control group (\textsf{Arm-1}), incoming query lists remain unchanged; in \textsf{Arm-k}, we always swap the document at the first position with the one at the $k$th position. The random swap assignment creates not only the intervention sets but also the unconfoundedness equation \eqref{eq: unconfoundedness}, so that we can compare the two methods without violating their own assumptions.

\subsubsection{Click Model}
\label{sec: 3.3.3}
Before generating clicks for query lists, we follow \citep{Joachims17} to binarize the $5$-grade relevance by setting $R = 1, \text{if } rel = 3, 4$, and $R = 0, \text{if } rel = 0, 1, 2$, and truncate lists to top-$10$. We then model the contextual examination by $P(E=1 | X=x, P=k) \coloneqq \frac{1}{k^{\mathrm{max}(w \cdot x + 1, 0)}}$ \citep{Fang19}, $w$ is drawn uniformly from $[-1, 1)$. To select the context features, we trained a random forests \citep{Breiman01} from the train and validation queries with normalized features and binary relevance, and we selected the top-$10$ important features as our context features\footnote{These context features encode the dependency between relevance and the context. Reducing the feature number in simulation also make the examination model more stable to produce enough variations in examination values.}. We also add click noise to the final click model by $P(C=1|E=1, R=1) = 1; P(C=1|E=1, R=0) = 0.1$ \citep{Joachims17, Jagerman19, Fang19}, modeling that a user can mistakenly click an irrelevant document after examining the document. 

\subsubsection{Position Bias Estimation}
To feed \textsf{CPBM} and \textsf{HTE Estimator} (i.e. Causal Forests and X-Learner), we process the simulation clicks in the two following ways:
\begin{itemize}
\item \textsf{CPBM}, merge clicks from all arms, select query-document pairs shown at least two positions (intervention sets) \citep{Agarwal19, Fang19}, and estimate the examination propensity using the method described in \ref{sec: 3.1}.
\item \textsf{HTE Estimator}, combine clicks between \textsf{Arm-1} and \textsf{Arm-k}, sample one position randomly from intervention sets for each unique query-document pair shown at position 1 and/or k, compute click rate at the sampled position, and estimate the heterogeneous treatment effect between position 1 and k with the method described in \ref{sec: 3.2}.
\end{itemize}
From the data process, we can see that not all the interventions are required for fitting HTE estimators. We train models on the simulation clicks generated from the train and validation sets of the query corpus. The training features are the same as the ones selected in the click model described in section \ref{sec: 3.3.3}. After this phase, we have now trained \textsf{CPBM} and \textsf{HTE Estimator}. These models will be used in the next phase in Figure~\ref{fig:ltr-diagram}.
\subsubsection{Unbiased Learning-to-rank}
Figure~\ref{fig:ltr-diagram} illustrates the architecture of unbiased LTR. We implemented two types of LTR models for the comparison. The main steps are:
\begin{itemize}
    \item \textsf{CPBM LTR}, estimate the examination propensity by \textsf{CPBM} and use the inverse propensity to weight the pairwise LTR.
    \item \textsf{HTE LTR}, compute click rate at the observed position, estimate the heterogeneous treatment effects with HTE estimators trained in the previous phase, compute the potential click rate at position 1 by the sum of the observed click rate and the treatment effect estimated, draw click/non-click by the $\mathrm{Bernoulli}$ distribution with the parameter of the potential click rate at position 1, and train the pairwise LTR.
\end{itemize}

The LTR models were implemented using the \emph{tensorflow-ranking} \citep{Pasumarthi19} library. We used a linear scoring function, pairwise hinge loss and L2 regularization. We used all features in the query corpus to train LTR models. Hyper-parameter tuning was conducted on the validation sets.

\subsubsection{Evaluation Metrics}
We evaluate the accuracy of the position bias estimation by computing the Root Mean Square Error (RMSE) between the estimated $\hat{\tau}_k(X_i)$ and the true $\tau_k^*(X_i)$ on queries in the test set for top-$10$ positions. \textsf{CPBM} does not output $\hat{\tau}_k(x)$ directly. Instead, it predicts the examination probability, $\hat{f_p}(x, k)$, and the average relevance, $\hat{f_r}(x, k, k')$, separately. We compute $\hat{\tau_k}(X_i)$ under \textsf{CPBM} by
$$\hat{\tau_k}(X_i)=\hat{f_p}(X_i, 1) \cdot \hat{f_r}(X_i, 1, k) - \hat{f_p}(X_i, k) \cdot \hat{f_r}(X_i, k, 1).$$
To evaluate ranking effectiveness we computed nDCG@10 on the test set and used binary relevance. We rerun 3 times of the entire simulation experiment, including simulation click generation, position bias estimations, and LTR training and evaluations. \footnote{Full experiment code is publicly available at \href{https://github.com/KimuraTian/sigir-eCom20-counterfactual-ltr-using-hte}{https://github.com/KimuraTian/sigir-eCom20-counterfactual-ltr-using-hte}.}

\section{Results}
In this section, we detail and analyze the experiment results comparing \textsf{CPBM} and the proposed methodology based on heterogeneous treatment effect estimation methods.
\begin{table*}[tbh]
\caption{Position bias estimations RMSE.}
\label{tbl:hte-estimation-rmse}
{\small
\begin{tabularx}{\textwidth}{ 
>{\raggedright\arraybackslash}X
>{\raggedleft\arraybackslash}X
>{\raggedleft\arraybackslash}X
>{\raggedleft\arraybackslash}X 
>{\raggedleft\arraybackslash}X 
>{\raggedleft\arraybackslash}X 
>{\raggedleft\arraybackslash}X 
>{\raggedleft\arraybackslash}X 
>{\raggedleft\arraybackslash}X 
>{\raggedleft\arraybackslash}X 
>{\raggedleft\arraybackslash}X 
>{\raggedleft\arraybackslash}X 
>{\raggedleft\arraybackslash}X 
>{\raggedleft\arraybackslash}X 
>{\raggedleft\arraybackslash}X 
>{\raggedleft\arraybackslash}X 
>{\raggedleft\arraybackslash}X 
>{\raggedleft\arraybackslash}X
}
\multicolumn{2}{l}{} 
& \multicolumn{4}{c}{Avg. Searches/Query = 5} 
& \multicolumn{4}{c}{Avg. Searches/Query = 10}
& \multicolumn{4}{c}{Avg. Searches/Query = 25}
& \multicolumn{4}{c}{Avg. Searches/Query = 50}
\\
\multicolumn{2}{l}{Percentage of Training Queries}
& 1\% & 10\% & 50\% & 100\% 
& 1\% & 10\% & 50\% & 100\% 
& 1\% & 10\% & 50\% & 100\% 
& 1\% & 10\% & 50\% & 100\% 
\\
\toprule
\multicolumn{2}{l}{CPBM}
& 0.360 & \textbf{0.256} & \textbf{0.239} & \textbf{0.237} 
& 0.325 & \textbf{0.249} & \textbf{0.239} & \textbf{0.238}
& \textbf{0.304} & \textbf{0.246} & \textbf{0.238} & \textbf{0.237}
& \textbf{0.295} & \textbf{0.245} & \textbf{0.239} & \textbf{0.237} 
\\
\multicolumn{2}{l}{Causal Forests}
& \textbf{0.335} & 0.276 & 0.255 & 0.249 
& \textbf{0.317} & 0.270 & 0.251 & 0.246
& 0.310 & 0.266 & 0.250 & 0.245
& 0.307 & 0.265 & 0.249 & 0.244 
\\
\multicolumn{2}{l}{X-Learner} 
& 0.374 & 0.315 & 0.275 & 0.260 
& 0.350 & 0.302 & 0.265 & 0.253
& 0.341 & 0.291 & 0.260 & 0.249
& 0.341 & 0.286 & 0.256 & 0.247 
\\
\bottomrule
\end{tabularx}
}
\end{table*}

\begin{figure*}[tb]
\includegraphics[width=\textwidth]{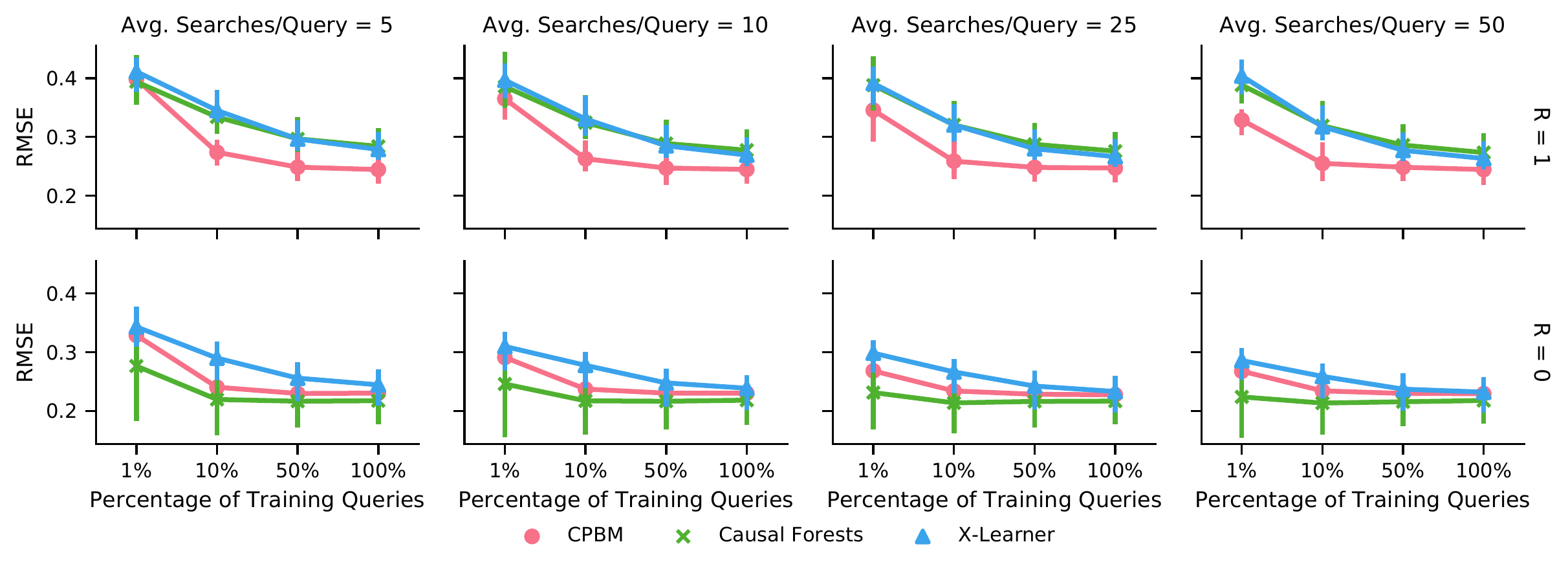}
\caption{Position bias estimations RMSE, as $\sqrt{\frac{1}{N}\sum_i\sum_k(\hat{\tau}_k(X^i) - \tau^*_k(X^i))^2}$. The columns and rows are the average searches per query and the true relevance of documents, respectively. X-axis represents the percentage of training queries used in the intervention simulation.}
\label{fig:res-bias-estimation}
\end{figure*}
\subsection{Position Bias Estimation}
Table~\ref{tbl:hte-estimation-rmse} shows RMSE of heterogeneous treatment effect estimations on the test set; models with the best RMSE are in bold. The RMSE gives us a picture of how accurate the estimator is able to capture the heterogeneity of the position bias. Among the three estimation methods, \textsf{X-Learner} has the largest estimation error. This may be because \textsf{X-Learner} is often more efficient when there is imbalance between the treatment and control groups or structural assumptions on heterogeneous treatment effects \citep{Kunzel19}. However, our experiment has balanced treatment and control groups, and the structure of the true relevance is unknown. This may suggest that future work will consider the ratio of treatment to control group and base learners for specific applications when picking HTE estimators. Under the extremely sparse condition (the percentage of training queries = 1\% and avg. searches per query = 5), \textsf{causal forests} method exhibits the smallest estimation errors. This is to be expected, since \textsf{causal forests} can utilize the similarities in the context feature spaces, reducing the high sample variance faced by \textsf{CPBM} due to the lack of intervention clicks. As we increase the density of intervention data points, \textsf{CPBM} has the best estimation accuracy. We conjecture this is related to the fact that \textsf{causal forests} uses the average value of the data points in the leaf node to make predictions, which introduces prediction noise when there is higher feature heterogeneity in the leaf node. When a large amount of intervention clicks is available, all the three algorithms tend to converge to a comparable level of estimation accuracy.

Figure~\ref{fig:res-bias-estimation} shows RMSE of heterogeneous treatment effect estimations for true relevant ($R=1$) and irrelevant ($R=0$) documents. For relevant documents, CPBM achieves the best RMSE except for the extremely sparse condition (avg. searches/query = 5 and percentage of training queries = 1\%), while both HTE estimators are comparable to each other. For irrelevant documents, \textsf{Causal Forests} has the best estimation accuracy. As our simulation modeled click noise, heterogeneous treatment effects for irrelevant documents are probabilities of misclicks. Although a good estimator should accurately capture heterogeneous treatment effects for both relevant and irrelevant documents, we speculate that the ranking performance of HTE LTR is less sensitive to the estimation noise of irrelevant documents compared to IPS weighted LTR, when click noise is not high. We think this is also related to the fact that misclicked irrelevant documents contribute to a small proportion of the clicked documents (e.g. maximum 0.1 under our simulation). Furthermore, in our HTE LTR approach the debiased clicks are generated by resampling, while the IPS weighting may amplify misclicks. 

\subsection{Unbiased LTR}
\begin{figure*}[tb]
\includegraphics[width=\textwidth]{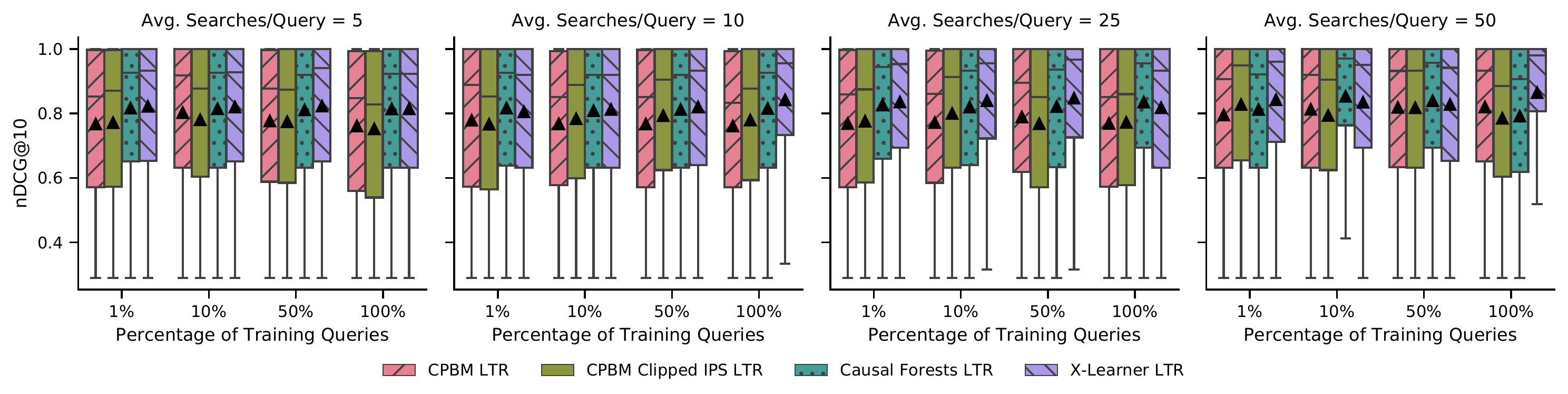}
\caption{Ranking performance of unbiased LTRs.}
\label{fig:res-ltr}
\end{figure*}

Figure~\ref{fig:res-ltr} shows the box plot of ranking accuracy results while Table~\ref{tbl:ranking-acc-ndcg10} shows ranking accuracy measured with nDCG@10. Overall, HTE LTR approaches (\textsf{Causal Forests LTR} and \textsf{X-Learner LTR}) outperform the IPS weighted LTR baseline (\textsf{CPBM LTR}). There are two possible reasons: \romannumeral1) noisy propensity estimations cause extremely high IPS when propensity scores are close to 0, which further explodes the weights of IPS weighted LTR loss function, making learning process unstable; \romannumeral2) the Bernoulli sampling can create positive clicks for documents without any observation clicks. To mitigate the high variance problem encountered by IPS weighted LTR, we trained a IPS weighted LTR with truncated propensity estimations in the range $[0.01, 1)$ (\textsf{CPBM Clipped IPS LTR}). \textsf{CPBM Clipped IPS LTR} improved the ranking performance over the baseline in some cases (e.g. avg. searches/query = 10 and percentage of training queries = 10\%, 50\%, 100\%, avg. searches/query = 25 and percentage of training queries = 1\%, 10\%), but it does not beat HTE LTR approaches. When clicks become abundant (i.e. avg. searches/query = 50), the performance of IPS weighted LTRs is significantly improved. For example, \textsf{CPBM LTR} beats \textsf{Causal Forests LTR} when 100\% of training queries are used for the propensity estimation. This may suggest that in applications where a large amount of intervention clicks is available, \textsf{CPBM LTR} is preferred as it requires simpler pre-processing. But when the lack of clicks is a main concern (e.g. personal file search, long tail queries), heterogeneous treatment effect estimation LTR methods may be better alternatives. Within the HTE LTR group, \textsf{Causal Forests LTR} and \textsf{X-Learner LTR} have similar ranking performance (cases when p > 0.05,\footnote{T-test with unequal variance on data from three runs.} the percentage of training queries = 1\%, 10\%, 100\% and avg. searches per query = 5, the percentage of training queries = 10\% and avg. searches/query = 10).
\begin{table*}[tb]
\caption{Ranking performance measured by nDCG@10. Notation $*$, $**$, and $***$ mean statistically significant with $p < 0.5$, $p < 0.01$, and $p < 0.001$, respectively, compared to \textsf{CPBM LTR}.}
\label{tbl:ranking-acc-ndcg10}
{\small
\begin{tabularx}{\textwidth}{ 
>{\raggedright\arraybackslash}X
>{\raggedright\arraybackslash}X
>{\raggedright\arraybackslash}X
>{\raggedright\arraybackslash}X 
>{\raggedright\arraybackslash}X 
>{\raggedright\arraybackslash}X 
>{\raggedright\arraybackslash}X 
>{\raggedright\arraybackslash}X 
>{\raggedright\arraybackslash}X 
>{\raggedright\arraybackslash}X 
>{\raggedright\arraybackslash}X 
>{\raggedright\arraybackslash}X 
>{\raggedright\arraybackslash}X 
>{\raggedright\arraybackslash}X 
>{\raggedright\arraybackslash}X 
>{\raggedright\arraybackslash}X 
>{\raggedright\arraybackslash}X 
>{\raggedright\arraybackslash}X
}
\multicolumn{2}{l}{} 
& \multicolumn{4}{c}{Avg. Searches/Query = 5} 
& \multicolumn{4}{c}{Avg. Searches/Query = 10}
& \multicolumn{4}{c}{Avg. Searches/Query = 25}
& \multicolumn{4}{c}{Avg. Searches/Query = 50}
\\
\multicolumn{2}{l}{Percentage of Training Queries}
& 1\% & 10\% & 50\% & 100\% 
& 1\% & 10\% & 50\% & 100\% 
& 1\% & 10\% & 50\% & 100\% 
& 1\% & 10\% & 50\% & 100\% 
\\
\toprule
\multicolumn{2}{l}{CPBM LTR}
& 0.77 & 0.80 & 0.78 & 0.76 
& 0.78 & 0.77 & 0.77 & 0.76 
& 0.77 & 0.77 & 0.79 & 0.77 
& 0.80 & 0.81 & 0.82 & 0.82 
\\
\multicolumn{2}{l}{CPBM Clipped IPS LTR}
& 0.77 & 0.78*** & 0.78 & 0.75** 
& 0.77*** & 0.78*** & 0.79*** & 0.78*** 
& 0.78* & 0.80*** & 0.77*** & 0.77 
& 0.83*** & 0.79*** & 0.82 & 0.79***
\\
\multicolumn{2}{l}{Causal Forests LTR}
& 0.82*** & 0.81*** & 0.81*** & 0.81*** 
& \textbf{0.82}*** & 0.81*** & 0.81*** & 0.82*** 
& 0.83*** & 0.82*** & 0.82*** & \textbf{0.84}*** 
& 0.81*** & \textbf{0.85}*** & \textbf{0.84}*** & 0.79***
\\
\multicolumn{2}{l}{X-Learner LTR} 
& \textbf{0.82}*** & \textbf{0.82}*** & \textbf{0.82}*** & \textbf{0.81}*** 
& 0.81*** & \textbf{0.81}*** & \textbf{0.82}*** & \textbf{0.84}*** 
& \textbf{0.84}*** & \textbf{0.84}*** & \textbf{0.85}*** & 0.82*** 
& \textbf{0.84}*** & 0.83*** & 0.83** & \textbf{0.86}***
\\
\bottomrule
\end{tabularx}
}
\end{table*}

\begin{figure*}[tb]
\includegraphics[width=\textwidth]{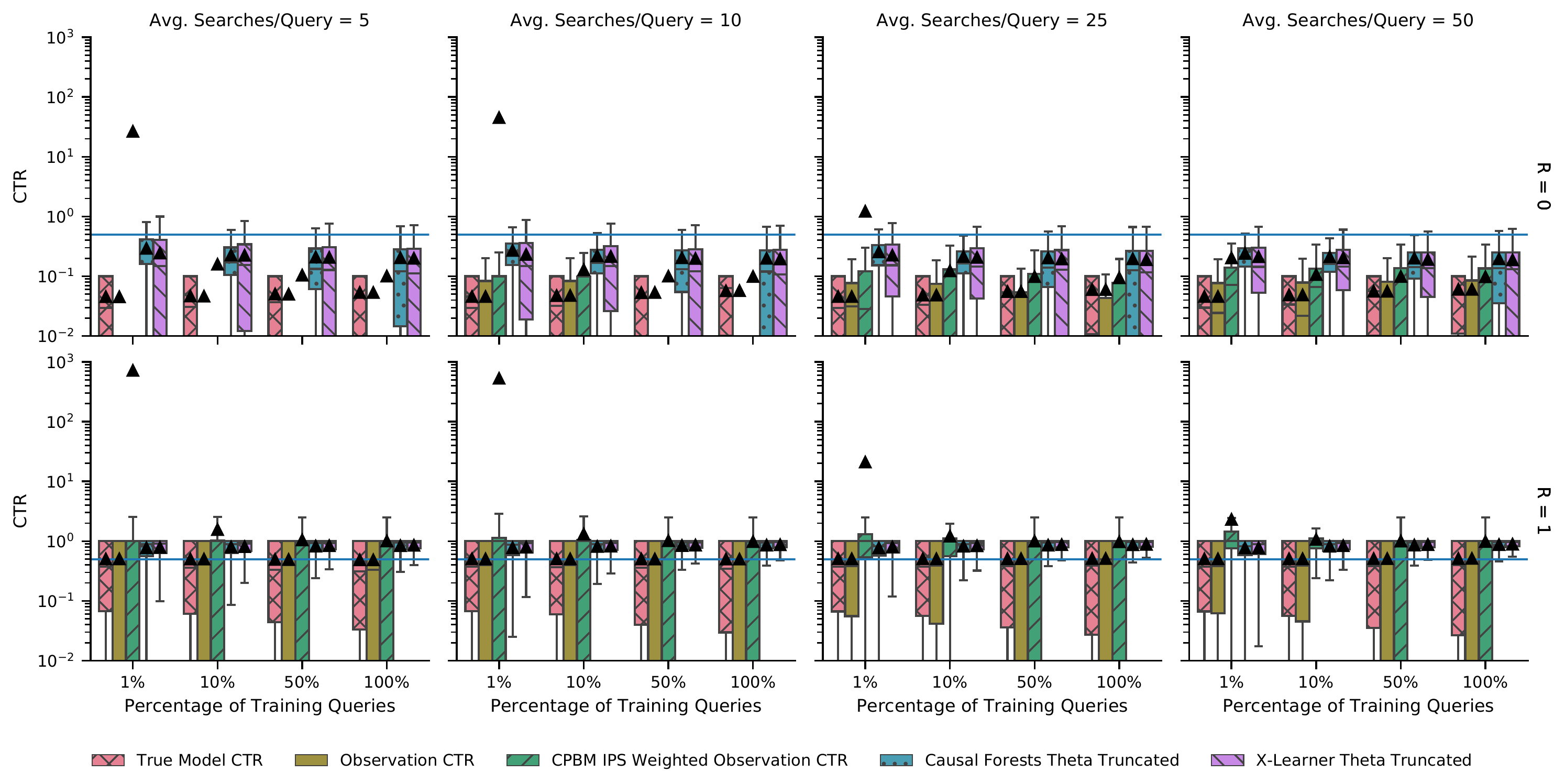}
\caption{Distribution of click rate in the training data.}
\label{fig:ctr-dist}
\end{figure*}

\begin{figure*}[tb]
\includegraphics[width=\textwidth]{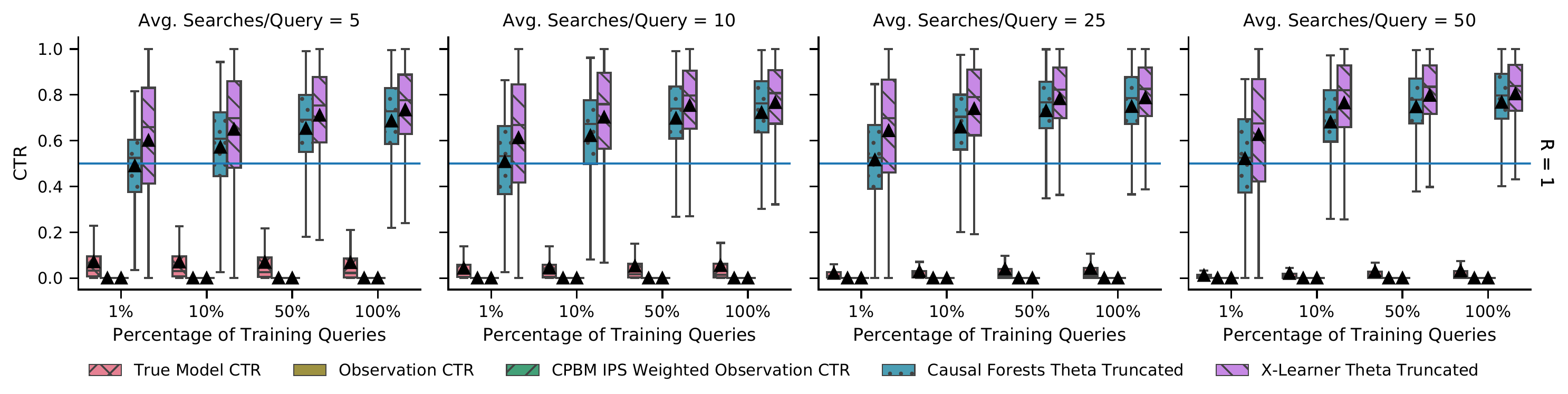}
\caption{Distribution of click rate for unobserved true relevant documents.}
\label{fig:ctr-dist-false-negative}
\end{figure*}

\begin{figure*}[tb]
\includegraphics[width=\textwidth]{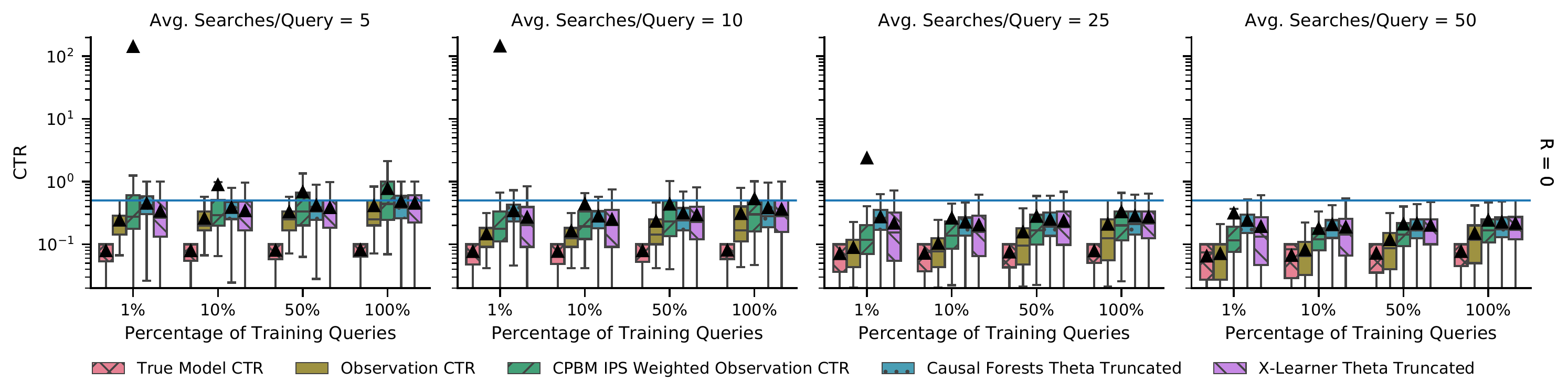}
\caption{Distribution of click rate for mis-clicked true irrelevant documents.}
\label{fig:ctr-dist-false-positive}
\end{figure*}
\subsection{Position Bias Results Analysis}
The core idea of counterfactual LTR is to estimate the bias in observational data to debias the training data and learn a better LTR model. To better understand the impact of the bias estimation on the ranking performance, we look at the distribution of corrected CTR and observed CTR in the training data. Figure~\ref{fig:ctr-dist} illustrates distribution of click rate for relevant documents and irrelevant documents. \textsf{True Model CTR} is the click probability defined in our click model (encoded with position bias); \textsf{Observation CTR} is observation CTR computed from simulation clicks; \textsf{Causal Forests Theta Truncated} and \textsf{X-Leaner Theta Truncated} are corrected CTR ($\theta$ of the Bernoulli sampling); \textsf{CPBM IPS Weighted Observation CTR} is the multiplication of IPS estimated by \textsf{CPBM} and observation CTR; the horizontal line marks $CTR=0.5$. For relevant documents, we expect corrected CTR is as close as to 1 since these are true relevant. For irrelevant documents, we expect corrected CTR to be as low as 0 and not over-estimate the false CTR. By looking at the bottom part of Figure~\ref{fig:ctr-dist}, we can see that the observation CTR almost covers the range $[0, 1]$ (high position bias), but it approaches to 1 after CTR adjustments. HTE LTR methods exhibit this property for all simulation conditions, while IPS weighted LTR only works well when there are more clicks available (small variance when avg. searches/query = 25, 50). On the other hand (top part of Figure~\ref{fig:ctr-dist}), we can see that HTE estimators, while having larger estimation variance on irrelevant documents, can still confine the corrected CTR of the majority of irrelevant documents under a certain level (e.g. 0.5), making the noisy estimation less detrimental to LTR. As long as relevant documents get sampled much more often than irrelevant documents, pairwise LTR can still differentiate relevant documents with irrelevant ones effectively. 

We also investigate how counterfactual LTR helps with debiasing observational data. Specifically, we focus on two cases: False Negative (Relevant documents without clicks) and False Positive (Misclicks) in observational data. Figure~\ref{fig:ctr-dist-false-negative} and Figure~\ref{fig:ctr-dist-false-positive} show the distributions. From Figure~\ref{fig:ctr-dist-false-negative}, HTE LTR recovers the majority of unobserved relevant documents through the click distribution correction. However, IPS weighted LTR does not help much, since it only weights observed clicks (always 0 for unobserved ones). For the false positive case (Figure~\ref{fig:ctr-dist-false-positive}), large estimation variance (e.g. extremely high means when avg.searches/query <=25 and percentage of training queries = 1\%) could be detrimental to IPS weighted LTR as it may severely amplify the noise.
\section{Conclusions and Future Work}
In this work we show how heterogeneous treatment effect estimation techniques can be used to address the position bias in search results ranking. To utilize estimated incremental causal effects for unbiased LTR, we drew clicks from debiased click distributions of observational data. We compared the proposed method with an existing contextual position-based model \citep{Fang19} under varying simulation conditions. Our results showed that the heterogeneous treatment effect estimation method is particularly effective for long tail queries with high click sparsity.

The usage of sampled clicks is not limited to pairwise LTR. It would be interesting to see how it performs when applied to other LTR methods. There is a variety of estimation methods for heterogeneous treatment effect, such as T-learner \citep{Kunzel19}, R-learner \citep{nie17}. In the future, we can explore the multivariate extension of R-learner \citep{nie17} so that we only need to build a single model for estimating bias in multiple positions instead of multiple models with the binary treatment indicator. The treatment position we chose in this work is the first position, we conjecture that the estimation could be improved if we extend to other anchor positions when there is huge imbalance of clicks across different positions. The simulation of user search behavior we used in this work (uniform sampling) is simple, future work may capture complex user behavior using other sampling distribution (e.g. Pareto distribution).

\begin{acks}
We thank Tao Ye, Jenny Lin, Oliver Bembom, Filip Korzeniowski, Ali Goli, Oscar Celma, and the Pandora Science Team, for their support.
\end{acks}

\bibliographystyle{ACM-Reference-Format}
\bibliography{counterfactual-ltr}


\end{document}